\documentclass[conference]{IEEEtran}
\pdfoutput=1
\newcommand{\mytitle}[0]{Distributed Triangle Counting \\ in the Graphulo Matrix Math Library}
\usepackage[  
  bookmarks=false,
  hyperfootnotes=false,
  hyperindex=false,
  hidelinks,  
  pdftitle={Distributed Triangle Counting in the Graphulo Matrix Math Library},  
  pdfauthor={Dylan Hutchison}]{hyperref}
  
\usepackage{cite}
\usepackage{url}
\usepackage{tikz}
\usepackage{footnote}
\usepackage{balance}

\usepackage[ampersand]{easylist}

\graphicspath{{.}{./img/}}
\usepackage{epstopdf}
\DeclareGraphicsExtensions{.eps,.pdf,.png}
\usepackage[cmex10]{amsmath}
\usepackage{todonotes}

\usepackage{dblfloatfix} 

\usepackage{graphicx}
\usepackage{subcaption}
\usepackage{amsfonts} 
\usepackage{amssymb} 
\usepackage{authblk}

\newcommand{\matr}[1]{\ensuremath{\mathbf{#1}}} 
\newcommand{\tr}[0]{{\mathsf{T}}} 

\usepackage{mathtools}

\usepackage{siunitx}
\usepackage{tabulary}

\usepackage{multirow}
\usepackage{adjustbox}
\usepackage{array}
\newcolumntype{R}[2]{%
    >{\adjustbox{angle=#1,lap=\width-(#2)}\bgroup}%
    l%
    <{\egroup}%
}

\makeatletter
\newcommand{\removelatexerror}{\let\@latex@error\@gobble}
\makeatother

\newlength{\algspace}


\usepackage{listings}

\usepackage[linesnumbered,lined]{algorithm2e}

\usepackage{color} 
\definecolor{mygreen}{RGB}{28,172,0} 
\definecolor{mylilas}{RGB}{170,55,241}

\begin{document}

\title{\mytitle{}}


\author[D. Hutchison] 
      {{\large Dylan Hutchison}\\
      	University of Washington}

%

\maketitle


\setcounter{footnote}{0}
\begin{abstract}
Triangle counting is a key algorithm for large graph analysis.
The Graphulo library provides a framework for implementing graph algorithms on the Apache Accumulo distributed database.
In this work we adapt two algorithms for counting triangles,
one that uses the adjacency matrix and another that also uses the incidence matrix,
to the Graphulo library for server-side processing inside Accumulo.
Cloud-based experiments show a similar performance profile for these different approaches 
on the family of power law Graph500 graphs, for which data skew increasingly bottlenecks.
These results motivate the design of skew-aware hybrid algorithms that we propose for future work.
\end{abstract}

\IEEEpeerreviewmaketitle


\section{Introduction}
\label{sIntro}

Today's data analytics continue to push the envelope in data size and complexity.
A class of NoSQL databases based on the Google Bigtable design \cite{chang2008bigtable} offers one solution framework: purchase as many commodity machines as needed, and stitch them together into a database cluster that provides performance on top of a bare-bones yet flexible data model that a user can adapt to particular applications.
Simple queries such as insert and scan perform well in these frameworks;
complex queries such as graph algorithms are difficult to implement in a way that realizes the performance capabilities of the database.

In this work we show a high performance implementation of the Static Graph Challenge \cite{gadepally2017static} on the Apache Accumulo distributed database within the Bigtable family.
Specifically we build on the graph processing abstractions provided by Graphulo, a matrix math library for Accumulo tables \cite{gadepally2015gabb}. 
Past work on Graphulo has focused on scaling up \cite{weale2016benchmarkinggraphulo} matrix multiplication \cite{hutchison2015graphulo} as well as the other GraphBLAS matrix math operations \cite{bader2014graph} and sample algorithms, including the k-Truss algorithm that is part of the Static Graph Challenge \cite{hutchison2016graphuloalg}.
We therefore focus on the remaining algorithm: triangle counting.
Triangles are defined as length-3 paths from a node to itself;
their frequency has many applications ranging from social network mining and cybersecurity to functional biology and link recommendation \cite{pavan2013counting}.

Specifically, we focus on counting triangles in large graphs that exceed main memory.
Accumulo stores large graphs on disk (in the Hadoop distributed file system); 
users expand disk and parallel compute capacity by adding more machines.
Burkhardt and Waring demonstrated the performance potential of large scale graph processing with the Accumulo database via a 70 trillion edge (scale 42, 1.1 PB) graph breadth-first search on a cluster of 1200 machines (57.6 TB collective memory) \cite{burkhardt2015cloud}.
Main-memory databases, on the other hand, require supercomputer-sized investments in order to process large graphs.
For example, the highest-scale Graph500 benchmark (based on \cite{bader2006designing}) submission as of June 2017 conducted breadth-first search on a 32 trillion edge graph (scale 41) with a national supercomputer consisting of 98k machines and over 1.5 PB collective memory.

We assume some familiarity with the Accumulo data model and the Graphulo method for in-database computation.
Accumulo's primary computational primitive is the \emph{range scan} over sorted and partitioned ranges of key-value entries that pass them through a series of \emph{server-side iterators}. 
Graphulo repurposes Accumulo's server-side iterators to read from additional tables and write to tables inside range scans.

The rest of the paper is organized as follows.
Section \ref{sAlgorithms} presents the mathematics and implementation of two methods for counting triangles with Graphulo. Both methods admit a number of interesting optimizations, and so it is unclear which one will scale better a priori.
Section \ref{sExperiments}'s experiments show that the two methods have similar performance profiles when run on power law data sets. We discuss a possible explanation in that the two methods face a common bottleneck---data skew---which motivates the design of a skew-aware hybrid algorithm we propose for future research.
Section \ref{sConclusion} concludes and comments on the potential of code generation for scaling out graph algorithm programmability to a wider audience.

\section{Algorithms} \label{sAlgorithms}
In this section we describe two algorithms for counting the number of triangles in a graph.
The first algorithm uses the graph's adjacency matrix as input;
the second algorithm uses both the graph's adjacency and incidence matrices as input.
Both algorithms require undirected graphs without self-edges.

\subsection{Adjacency-only Triangle Counting}

\begin{algorithm}[t]
  \DontPrintSemicolon
  \SetCommentSty{textit}
  \KwIn{Unweighted adjacency matrix $\matr A$}
  \KwOut{Number of triangles $t$}
  \begingroup
  \def\B{\matr{B}}
  \def\C{\matr{C}}
  \def\L{\matr{L}}
  \def\U{\matr{U}}
  \def\A{\matr{A}}
  Split $\A$ into $\L + \U$ \tcp*{lower and upper triangle of \A} 
  $\B = \L \U$              \tcp*{matrix multiply}
  $\C = \B * \A$        \tcp*{element-wise multiply (mask) with \A}
  $t = \operatorname{sum}(\C)/2$ 
  \endgroup
\vspace{5pt}
\caption{Cohen's triangle counting}
\label{algCohen}
\end{algorithm}

Our first algorithm adapts Cohen's algorithm \cite{cohen2009graph} in order to run in two passes.
We briefly review Cohen's algorithm in Algorithm \ref{algCohen}.
Cohen's algorithm first computes the set of wedges (paths of length 2) by multiplying the lower and upper triangles of the adjacency matrix \matr{A}.
It then restricts these wedges to the set of wedges that are closed by masking the result with \matr{A}, which is an element-wise operation. Closed wedges are triangles.
The number of triangles is given by counting the number of closed wedges and dividing by two, since each wedge is formed twice in the matrix multiply.

\begin{algorithm}[t]
  \DontPrintSemicolon
  \SetCommentSty{textit}
  \KwIn{Upper triangle of unweighted adjacency matrix $\matr A$}
  \KwOut{Number of triangles $t$}
  \begingroup
  \def\T{\matr{T}}
  \def\A{\matr{A}}
  $\T = \A$ \tcp*{clone $\A$ to $\T$}
  $\T = \T + \operatorname{triu}(\matr{A}^\tr \A)$ \tcp*{upper triangle of matrix multiply}
  \tcp*{custom multiply: $a \otimes b = 2$ if $a=b=1$, otherwise 0}
  $\T(\T \operatorname{\%} 2 == 0) = 0$ \tcp*{filter to odd entries}
  $t = \operatorname{sum}((\T-1)/2)$ 
  \endgroup
\vspace{5pt}
\caption{Graphulo Adjacency-only triangle counting}
\label{algAdj}
\end{algorithm}

The Graphulo adaptation of Cohen's algorithm is given in Algorithm \ref{algAdj}.
We made the following changes to Cohen's algorithm in order to reduce the number of intermediate entries produced as well as the number of times each entry is read:
\begin{enumerate}
\item Only use the upper triangle of the adjacency matrix by rewriting the matrix multiply $\matr{L}\matr{U}$ as $\matr{U}^\tr\matr{U}$.
This form admits the one-pass outer product matrix multiply algorithm \cite{hutchison2015graphulo}
and also cuts the input in half.
It also cuts the output of $\matr C$ in half, which is what we would do in line 4 of Cohen's algorithm anyway.
\item Filter the output of the matrix multiply to the upper triangle.
Because the lower triangle output of the matrix multiply will be zeroed during the element-wise multiply anyway, the lower triangle can be pruned early, before writing to $\matr T$.
\item Add the result of the matrix multiply into $\matr A$ via an Accumulo table clone and a ``parity trick'' to determine when matrix multiply entries overlap with $\matr A$.
Double each partial product from the matrix multiply, making them all even, which allows entries that overlap with $\matr A$ to be detected by checking for odd parity.
The parity trick eliminates the need to do a further element-wise operation with $\matr A$.
\end{enumerate}

The parity trick is one way of performing a \emph{masked matrix multiply} in the Accumulo database. In-memory databases can implement masks more directly by holding $\matr A$ in memory and restricting the output of the matrix multiply to those overlap with $\matr A$ right away. Because Accumulo is an out-of-core, distributed database, it cannot prune entries outside of $\matr A$ right away because it cannot hold $\matr A$ in memory, which is always true for large enough graphs. It could even be the case that partial products are written to separate files, which means that we cannot eagerly check for the presence of the 1 from $\matr A$. Instead we use the parity trick in a delayed fashion, filtering entries during a scan of \matr{T} after all partial products are written.

Given the mathematics in Algorithm \ref{algAdj}, we now describe its Graphulo implementation.
The row and column of matrix entries are stored in their string encoding in the row and column qualifier of Accumulo entries.
With this schema, Accumulo partitions $\matr A$ into a set of tablets, each of which contain a sorted, consecutive block of $\matr A$ at the granularity of rows; every row resides within a single tablet.
We choose the particular splits that partition the rows of $\matr A$ into tablets
such that each tablet contains an approximately equal share of $\matr A$'s entries.
Accumulo assigns these tablets evenly among all available tablet servers.\footnote{Tablet servers are the worker machines of an Accumulo database.}
Compacting \matr{A} ensures these splits take effect.

The table clone of $\matr A$ to $\matr T$ is a ``copy-on-write'' metadata operation that only results in new data files when data is written to $\matr T$. 
The cloned table $\matr T$ has the same splits as $\matr A$. 

The $\matr{T} = \matr{T} + \operatorname{triu}(\matr{A}^\tr \matr{A})$ is implemented as a fused Graphulo TableMult operation on \matr{A} with itself. The TableMult operation uses a custom ``row multiply'' function that applies the custom multiply function and filters the generated partial products to the upper triangle. These operations run within a scan of each of \matr{A}'s tablet servers and write their entries to \matr{T}. 

Specifically, the TableMult acts on entries from \matr{A} of the form $(r,c,1)$ and $(r,c',1)$ where $r$ is a row of $\matr A$, $c$ and $c'$ are columns of $\matr A$, and `1' is the value of the entries (since $\matr A$ is unweighted). 
The result of the TableMult are entries of the form $(c, c', 2)$ where $c < c'$, and these are written to \matr{T}.
At $\matr T$, partial products are summed together by standard server-side iterators during flushes (when Accumulo spills entries from memory to disk) and compactions (when Accumulo merges files on disk together).

When the matrix multiplication completes (after all partial products are written to $\matr T$),
a Reduce operation is initiated on $\matr T$.
For each value from the matrix multiply (fully summed together from the partial products),
we (1) filter the entries to only accept odd values,
(2) transform the value by $v = (v-1)/2$,
and (3) sum together all values.
This sum runs independently on each tablet of $\matr T$, resulting in partial sums which are collected at a client and summed together into a final count of the number of triangles, as per a standard user-defined aggregation pattern \cite{cohen2006user}.

\subsection{Adjacency+Incidence Triangle Counting}

\begin{algorithm}[t]
  \DontPrintSemicolon
  \SetCommentSty{textit}
  \KwIn{Lower triangle of unweighted adjacency matrix $\matr A$}
  \KwIn{Unweighted incidence matrix $\matr E$}
  \KwOut{Number of triangles $t$}
  \begingroup
  \def\E{\matr{E}}
  \def\T{\matr{T}}
  \def\A{\matr{A}}
  $\T = \operatorname{triu}(\matr{A}^\tr \E)$ \tcp*{upper triangle of matrix multiply}
  $t = \operatorname{sum}(\T == 2)$  \tcp*{count the entries of \T{} equal to 2}
  \endgroup
\vspace{5pt}
\caption{Graphulo Adj.+Incidence triangle counting}
\label{algAdjEdge}
\end{algorithm}

Our second algorithm uses both the adjacency and incidence matrix of a graph as input.
In contrast to the first algorithm, the second only uses the lower triangle of the adjacency matrix.
We define the incidence matrix as the matrix whose rows are vertices, whose columns are edges, and whose values, for vertex $v$ and edge $e$, are defined as $\matr{E}(v,e) = 1$ if $e$ is incident on $v$ and 0 otherwise.
This definition requires every column of \matr{E} to have exactly 2 nonzero entries.

Our algorithm takes inspiration from Wolf's triangle enumeration algorithm \cite{wolf2015task}.
Algorithm \ref{algAdjEdge} specializes Wolf's algorithm to triangle counting and a Graphulo implementation.

The algorithm identifies triangles by combining two pieces of information: that the presence of a 2 in the $\matr{A} \matr{E}$ indicates that one vertex has a connection to two other vertices, and that these two other vertices because they are connected in the incidence matrix.
Our adaptation restricts \matr{A} to its lower triangle and the output of the $\matr{A}^\tr \matr{E}$ to its upper triangle
in order to eliminate redundant computation.\footnote{In Wolf's algorithm, every triangle is enumerated three times and counting triangles requires division by 3 after summing entries, just as Cohen's algorithm requires division by 2 after summing entries.}

Because the incidence matrix is not square, 
the notion of ``upper triangle of $\matr{A}^\tr \matr{E}$'' requires further explanation.
Each vertex is encoded into the rows and columns of \matr{A} and the rows of \matr{E} as normal.
Each edge is encoded into the columns of \matr{E} as the concatenation of the vertex labels that the edge is incident on \emph{in ascending order}.
Thus, each edge is stored as the pair of vertices $[v_1, v_2]$, where $v_1 < v_2$.\footnote{$v_1 \neq v_2$ because there are no self-edges.}
We define the upper triangle of \matr{E} as the restriction of \matr{E} onto only the entries $(v, [v_1,v_2])$ where $v < v_1$.

We now describe Algorithm \ref{algAdjEdge}'s Graphulo implementation.
Like the parity trick in Algorithm \ref{algAdj}, we use data format tricks in order to compare vertices to edges and to distinguish counted triangles from lone partial products.

We switched from a string encoding to a fixed 4-byte encoding of the vertex labels\footnote{Four bytes per vertex is sufficient for this work. The number of bytes per label is not significant in general because Accumulo uses run-length encoding.}
in order to facilitate the concatenation and un-concatenation of vertices.
Thus, the columns of \matr{E} are stored as 8-byte labels composed of two 4-byte vertex labels. 

We split \matr{A} and \matr{E} on their rows into approximately equal sized tablets and compact them. We set the splits of intermediary \matr{T} to the same splits as \matr{E}.

We implemented the matrix multiply $\operatorname{triu}(\matr{A}^\tr \matr{E})$ as another fused Graphulo TableMult.
This TableMult eagerly filters its output to the upper triangle. 
Its output value is an empty (0-byte) value, which serves as a marker for one partial product of the matrix multiply.
Because the incidence matrix only has two nonzero values per column, only two partial products per entry are possible.
In total, on input $(v,v_1,1)$ from \matr{A} and $(v,[v_2,v_3],1)$ from \matr{E}, 
the TableMult writes the entry $(v_1, [v_2,v_3], $`'$)$ to \matr{T} when $v_1 < v_2$.

We run two special aggregation iterators during \matr{T}'s flushes and compactions in order to pre-sum the result of $\operatorname{sum}(\matr{T} == 2)$ while entries are being written to \matr{T} in the middle of the matrix multiply.
Pre-summing during the matrix multiply is important because it reduces the number of entries written to disk, reducing Accumulo's write bottleneck and speeding up the full $\operatorname{sum}$ reduction that takes place once the matrix multiply finishes.

The first aggregation iterator watches for two consecutive empty values that have the same key (in our case, the same row and column). When this condition occurs, it means that we have found an entry in \matr{T} whose partial products sum to 2, which indicates a triangle. The first iterator replaces these two entries with empty values with an entry with value 1, to indicate the triangle. This computes $\matr{T} == 2$.

The second aggregation iterator sums together values that are numbers (i.e., values that are not empty) irrespective of their keys. Entries with empty values pass through.
This iterator effectively performs early aggregation of discovered triangles, 
which can be done even before the $\operatorname{triu}(\matr{A}^\tr \matr{E})$ matrix multiply finishes.

When the matrix multiply does finish, we initiate a full $\operatorname{sum}$ reduction by scanning \matr{T}.
This last scan sums together non-empty values, which may already be partially summed as a result of the second aggregation iterator.
A client gathers the local sums from each tablet server and sums those into a final triangle count.

\section{Experiments} \label{sExperiments}
\subsection{Setup and Results}

We ran experiments testing the scalability of the two algorithms on a cloud deployment of Accumulo.
We present the experiment details first, followed by results and discussion.

We tested Graphulo's triangle counting algorithms on data from the synthetic Graph500 RMAT unpermuted power law graph generator \cite{leskovec2005realistic}.
We chose to run on power law data because it well models many real world applications \cite{gadepally2015using} while also being experimentally convenient, since the structure of the graph remains the same as graph size increases.
The generator creates matrices that range from $2^{10}$ rows (scale 10) to $2^{20}$ rows (scale 20), with roughly 16 times that many nonzero entries.
In order to create undirected adjacency matrices, we added the resulting matrix to its transpose, eliminated the diagonal, and set all nonzero values to 1.\footnote{To reproduce our power law graphs, download Octave 4.2.1 and D4M \cite{kepner2012dynamic}, set the random seed in Octave as \texttt{rand('seed',20160331)}, run the D4M file \texttt{KronGraph500NoPerm.m}, eliminate the diagonal, and add the result to its transpose.}

We deployed Accumulo onto an Amazon EC2 cluster of 12 \texttt{m3.xlarge} machines, 
consisting of 8 tablet servers, 3 coordinators (for Zookeeper, the Hadoop NameNode, and the Accumulo Master),
and a monitor machine that tracks the health of the others.
Each machine has two 40 GB SSDs and 4 vCPUs on a 2.5 GHz Intel Xeon E5-2670v2.
The cost of this cluster is \$3.192/hour, 
though only the 8 tablet server machines make up a variable cost component in terms of cluster scalability 
(the 3 coordinators and monitor are essentially fixed costs).
For each tablet server, we allocated 8 GB to tablet server Java heap memory, 2 GB to data cache, 1 GB to index cache, and 2 GB to Accumulo's native in-memory maps.
Amazon rates the network performance of these machines as ``high'' but does not guarantee a particular level of throughput or latency.
Our experiments are I/O-bound and therefore vulnerable to cloud-based network and disk performance variance \cite{zhai2011cloud},
but it is unclear how much variance actually affected our experiments.


We split each table into at most 24 tablets.  
This number of tablets appeared to provide the best performance during an initial experiment.

Our performance metrics are as follows:
\begin{itemize}
  \item \emph{runtime} is the best recorded time to count the number of triangles across 2 or 3 runs.
  \item \emph{nedges} is the number of edges, which is the number of nonzero entries in the upper (or lower) triangle of the adjacency matrix $\matr A$. 
  The structure of the incidence matrix gives nnz($\matr{E}$) = $2*$nedges.
  \item \emph{nppf} is the number of partial products formed as a result of the matrix multiply in each algorithm \emph{after} applying the upper triangle filter.\footnote{The total number of partial products is a bit more than double nppf.}
  It holds that nppf $>\!>$ nedges due to the nature of matrix multiply.
  The real workload of this task is therefore due to nppf.
\end{itemize}

Because each partial product is processed at least twice---once by a matrix multiply and once by a reduce---we calculate the processing rate as $2 * nppf / runtime$.
This rate more accurately captures the work required to count triangles with these algorithms than nedges / runtime, because the work required is quadratic in each vertex's degree and the input edge count hides this fact. We do not count filtered-out partial products because the computation is I/O-bound and the result of the matrix multiply is far larger than the input, and so filtered-out partial products are not as significant a factor.


\begin{figure*}[t]
\centering
\begin{minipage}[c]{1.0\columnwidth}
\includegraphics[width=\linewidth]{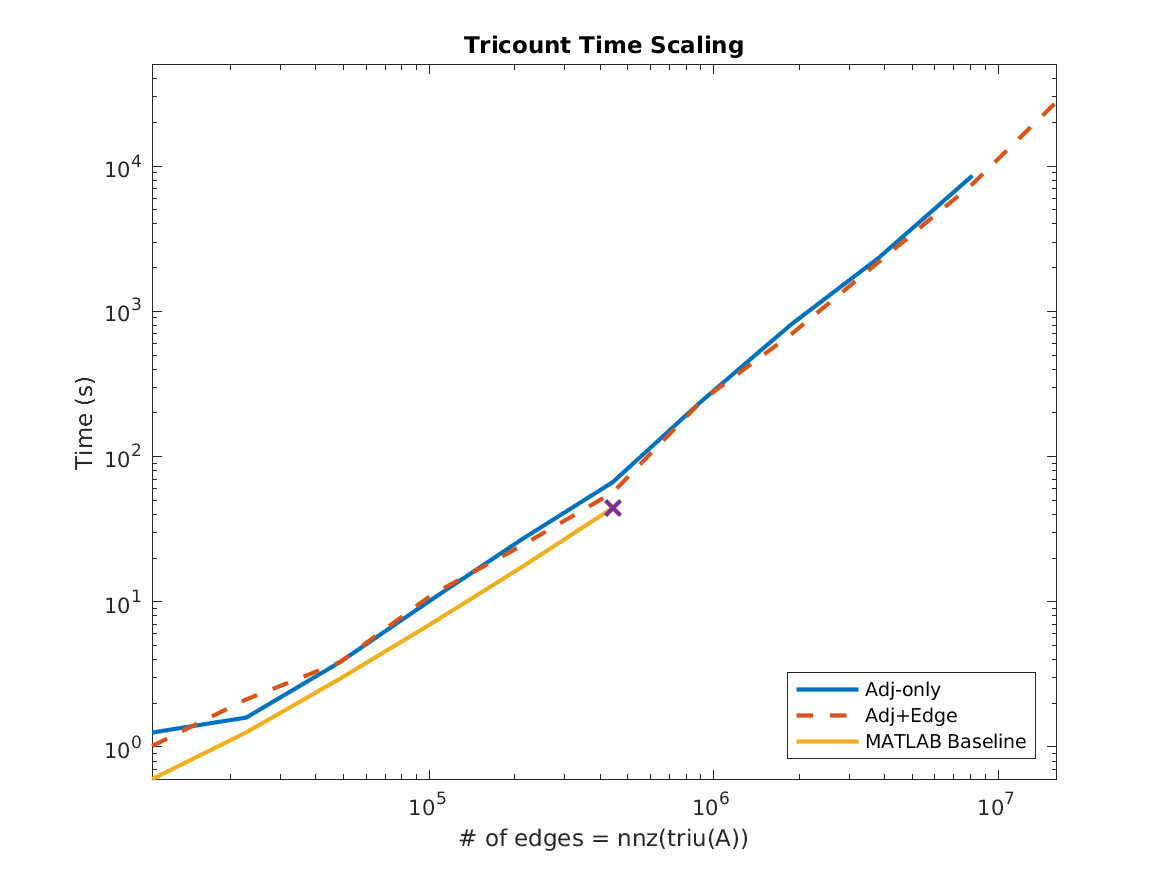}
\caption{Baseline and Graphulo triangle counting runtime.}
\label{fTriTime}
\end{minipage}
\hfil
\begin{minipage}[c]{1.0\columnwidth}
\includegraphics[width=\linewidth]{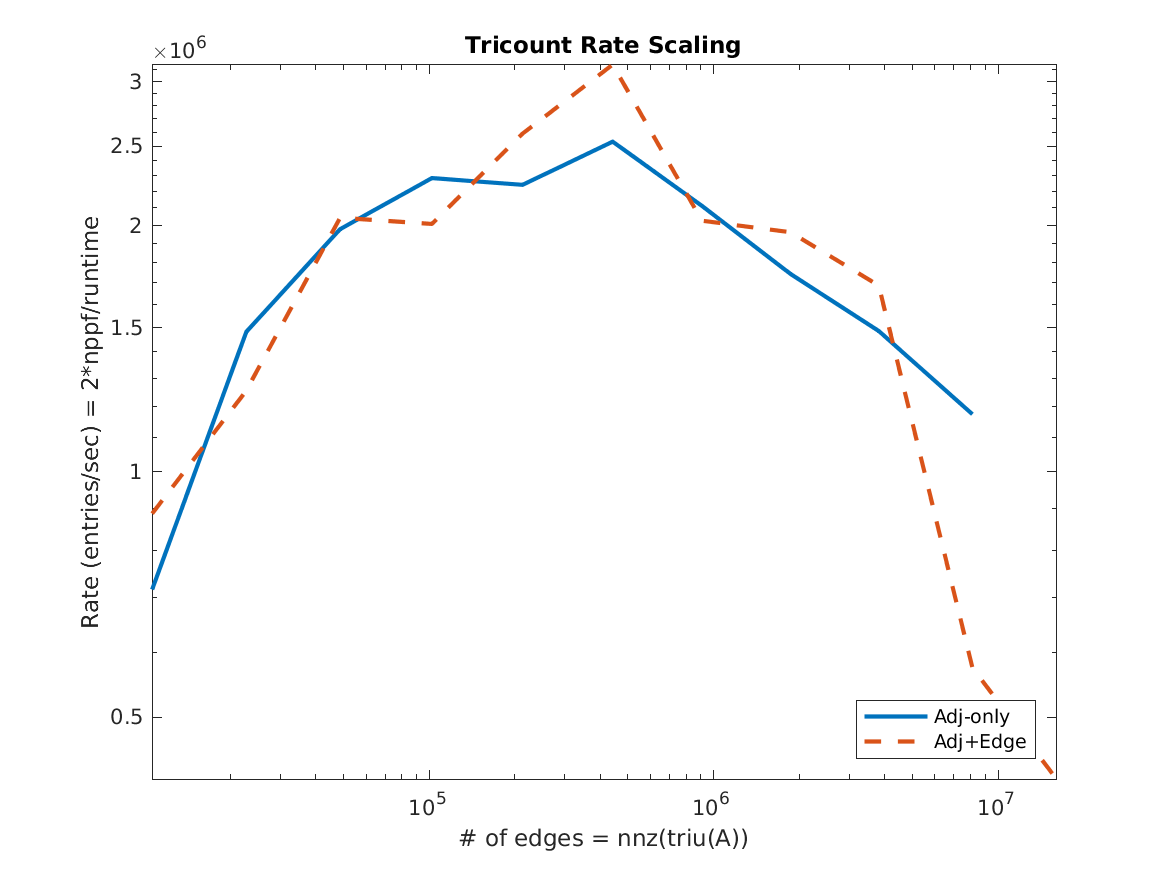}
\caption{Graphulo processing rate during triangle counting.}
\label{fTriRate}
\end{minipage}
\end{figure*}

\begin{table*}[tb]
\centering
\begin{tabular}{c|c|c|c|c|c|c|c|c}
\multirow{2}{1.75em}{\adjustbox{angle=30,lap=\width-3.75em}{SCALE}} &  & \multicolumn{3}{c|}{Adjacency-only Algorithm} & \multicolumn{3}{c|}{Adjacency+Incidence Algorithm} & Baseline \\
 & \# of edges & nppf (entries) & Time (s) & Rate (entries/s) & nppf (entries) & Time (s) & Rate (entries/s) & Time (s) \\
\hline
10 & \num{10619.000} & \num{450321.000} & \num{1.257} & \num{716501.193} & \num{450392.000} & \num{1.015} & \num{887384.494} & 0.60 \\
\hline
11 & \num{22758.000} & \num{1180321.000} & \num{1.594} & \num{1481140.670} & \num{1334328.000} & \num{2.126} & \num{1255070.310} & \num{1.261} \\
\hline
12 & \num{48616.000} & \num{3800155.000} & \num{3.842} & \num{1977959.662} & \num{3921738.000} & \num{3.840} & \num{2042678.264} & \num{2.954} \\
\hline
13 & \num{102239.000} & \num{11862521.000} & \num{10.386} & \num{2284329.097} & \num{11233750.000} & \num{11.193} & \num{2007245.470} & \num{7.106} \\
\hline
14 & \num{212859.000} & \num{30352452.000} & \num{27.087} & \num{2241100.155} & \num{31736154.000} & \num{24.516} & \num{2588973.422} & \num{17.449} \\
\hline
15 & \num{441810.000} & \num{84227504.000} & \num{66.555} & \num{2531057.047} & \num{88821778.000} & \num{56.481} & \num{3145169.135} & \num{44.259} \\
\hline
16 & \num{909555.000} & \num{255500322.000} & \num{241.900} & \num{2112442.328} & \num{245726068.000} & \num{242.612} & \num{2025670.344} &  \\
\hline
17 & \num{1864858.000} & \num{702935396.000} & \num{806.489} & \num{1743199.832} & \num{676746631.000} & \num{690.327} & \num{1960654.406} &  \\
\hline
18 & \num{3805915.000} & \num{1734071173.000} & \num{2335.956} & \num{1484678.259} & \num{1842786827.000} & \num{2183.723} & \num{1687747.707} &  \\
\hline
19 & \num{8126523.000} & \num{5044863964.000} & \num{8592.616} & \num{1174232.393} & \num{2159538720.000} & \num{7513.619} & \num{574833.187} &  \\
\hline
20 & \num{16057587.000} &  &  &  & \num{5820403494.000} & \num{27724.255} & \num{419878.080} &  \\
\end{tabular}
\caption{Tricount algorithm experiment metrics. nppf = number of partial products after filtering to upper triangle.}
\label{tTricountTable}
\end{table*}

Figure \ref{fTriTime} plots the runtime of the two triangle counting algorithms 
on a log-log scale.
The two algorithms show very similar performance profiles within an order of magnitude.
Table \ref{tTricountTable} tabulates the results for closer inspection.
Figure \ref{fTriRate} plots the processing rate defined above.


\subsection{Comparison to Graph Challenge Baseline}
The Graph Challenge baseline implementations are designed for a very different execution environment and problem size than that of Graphulo.
When the input graph (and intermediary results) fit into the memory of a single node, we expect that the baseline implementations, all of which are in-memory, will count triangles faster than Graphulo, even without optimizing them with algorithmic tricks or multi-threading.
However, the baseline implementations cannot scale to large graph sizes that Graphulo can handle.

We ran the MATLAB baseline implementation\footnote{The MATLAB baseline computes $t = \operatorname{nnz}(\matr{A}\matr{E} == 2)/3$. The intermediary result \matr{A}\matr{E} is the memory bottleneck.} on one machine from the Amazon cluster, without any other processes running, and found that the implementation exceeded the machine's 15GB of memory at graphs larger than scale 15 (524k edges). 

We plot the baseline runtime up to scale 15 alongside the Graphulo runtime in Figure \ref{fTriTime}
and in Table \ref{tTricountTable}.
The baseline runtime include the time to load data from a file into an adjacency and incidence matrix.

While the comparison of Graphulo with a baseline in-memory implementation is a good indicator of triangle counting performance, we caution users on making infrastructure decisions solely on the basis of this comparison.
Users usually have additional requirements beyond triangle counting, such as indexed access to graph subsets and the ability to apply custom filters.
These additional, holistic requirements often favor storage in a database, such as Accumulo or other graph systems, rather than storage in flat files as in the baseline implementation.

\subsection{Discussion: Skew \& Hybrid Matrix Multiply} 
\label{sDisc}
Several observations led us to diagnose skew in the graph's degree distributions (i.e., the presence of high-degree vertices) as a major problem.
First we noticed that the choice of row and column encoding---whether to encode the rows as 4-byte fixed-width integers vs. a UTF-8 string encoding---impacted runtime significantly.
The adjacency-only algorithm ran 2x faster under the string encoding. 
We believe that the change of format led to a permutation on the rows and columns that affects the number of partial products computed at each row of \matr{A} (due to a permuted ordering) and therefore load balance.
Past work on sparse matrix multiplication has likewise noted that permuting the input matrix can reduce runtime \cite{azad2016exploiting}, and so we expect better performance, to a certain extent, if we explicitly permute the generated input matrices.

We also noticed that the computation's bottleneck shifts as graph size increases.
The reduction operation bottlenecks at lower graph sizes, whereas the matrix multiply operation bottlenecks at larger graph sizes (and increasingly more so as graph size increases). 
A phase transition occurs between scales 15 and 16 (between 520K and 1.05M edges),
at which point the time to compute the matrix multiply exceeds the time to compute the reduction.
Interestingly, this is also the graph size range at which both algorithms achieve peak processing rate.

We attribute the shift of bottleneck to the power law nature of the input graphs, and in particular a few high-degree vertices. Because matrix multiply is quadratic in the degree of the each vertex, the presence of high-degree vertices leads to skew in the matrix multiply work load among the tablet servers. 

We similarly observed these load balancing problems during the matrix multiply (and to a lesser extent during the reduction), in which one tablet server takes far longer to finish than the others.
This skew is unavoidable in the sense that, no matter how the rows of $\matr A$ are partitioned among the tablet servers, some tablet server must have the highest-degree vertex and will take far longer to process that than the other tablet servers.
Thus, even though we could have chosen to split \matr{A} and \matr{T} in a non-uniform manner that is application-specific and possibly even data-dependent (which may preclude their use on general tables), a side experiment showed that these changes skirt the main problem of high-degree nodes.

Both the database literature \cite{beame2014skew} and the high performance computing literature \cite{ballard2016hypergraph} have studied the problem of skew in detail. We recommend adapting some of their techniques to Graphulo by, for example, inserting the first bits of the column qualifier into the column family and leveraging locality groups in order to implement a 2-D partitioning strategy that spreads high-degree vertices among the tablet servers. This strategy could be applied to every vertex, as in Sparse SUMMA \cite{buluc2008challenges}, or just to the high-degree vertices in a degree-aware manner, as in Hypercube Join \cite{chu2015theory}.

One way we might solve the problem of high-degree vertices is by reconsidering the 1-D inner product matrix multiply algorithm. In general, inner product matrix multiply is extremely inefficient on the Accumulo database because, for $\matr{C} = \matr{A}\matr{B}$, the matrix \matr{B} must be read $n$ times in full, where $n$ is the number of rows of \matr{A} (or vice versa). This is prohibitive for even moderately sized matrices, since these entail disk reads.

However for this particular matrix multiply, $\matr{T} = \matr{A}^\tr \matr{A}$, the inner product algorithm has several advantages.
First, only the upper (or lower) triangle is required, which eliminates half the times that $\matr{A}$ must be read.
More importantly, at the time the entry $\matr{T}(r,c)$ would be computed, the corresponding row $\matr{A}^\tr(r,*)$ and column $\matr{A}(*,c)$ are held in memory. The computation can be avoided if $\matr{A}(r,c) = 0$ because the resulting entry would be masked by the future element-wise multiply.
Last, the inner product algorithm fully sums together the partial products of each entry in the result $\matr{T}$. These entries do not need to be materialized because the reduce operation applies immediately; the triangle count can be computed during the matrix multiply.

The hybrid algorithm we propose is to run inner product on high-degree vertices and outer product on all other vertices.
Whereas outer product is expensive on high-degree vertices, inner product handles them efficiently by immediately summing their partial products and avoiding the writing of values that would be zeroed by the element-wise multiply mask.
Whereas inner product is too expensive to run on every vertex, running inner product just on the high-degree vertices is reasonable.

In general, our proposed hybrid algorithm exploits the idea that the optimal choice of an operator's implementation depends on its context.
In databases, a sort-merge join may outperform an otherwise optimal hash join in databases when followed by a range selection on that same sort order \cite{selinger1979access}. 
In high performance computing, the optimal choice of matrix multiplication algorithm changes when the matrix multiply is considered in the context of a loop, rather than in isolation \cite{koanantakool2016communication}.
The same concept may apply here as well, in the sense that an inner product matrix multiply may outperform an otherwise optimal outer product matrix multiply when followed by a mask and full aggregation, as in the triangle counting problem.

\section{Conclusion} \label{sConclusion}
In this work we adapted two algorithms for triangle counting,
one that uses on the adjacency matrix and another that uses the adjacency and incidence matrix,
to the Graphulo library for server-side processing on the Apache Accumulo database.
Experiments show a similar performance profile for these different approaches 
on power law synthetic graphs.

In future work we recommend investigating hybrid algorithms that better handle high-degree vertices, such as the one proposed in Section \ref{sDisc}.
We also recommend studying real-world graphs that likely exhibit less skew than the Graph500 power law synthetic graphs whose high levels of skew are unrealistic for many applications.

An alternative strategy is exporting data from Accumulo to an external system and running a specialized algorithm there, such as parallel shared-memory triangle counting \cite{shun2015multicore}.
Polystore systems such as Myria \cite{wang2016myriaOverview}, BigDAWG \cite{gadepally2016bigdawg}, and Rheem \cite{agrawal2016rheem} facilitate using multiple systems together via data movement techniques, as in PipeGen \cite{haynes2016pipegen} and Portage \cite{dziedzic2016data}.

User programmability is another important area in the sense of ``how easily can users write custom, application-specific graph algorithms without sacrificing performance?'' \emph{Code generation} is a modern technique often used (e.g., in SystemML \cite{boehm2014systemml}) to bridge the gap between higher-level APIs that are easily programmable and lower-level code that implement efficient data structures and runtime tricks, such as those employed for counting triangles in this work. 
The LaraDB system \cite{hutchison2017laradb} prototypes this approach by compiling programs in the high-level Lara algebra to Graphulo iterator code, and we expect further future work to push this approach.

\section*{Acknowledgment}
This material is supported in part by NSF Graduate Research Fellowship DGE-1256082.
Cloud computing was provided by the University of Washington Student Technology Fee Grant 2017-96
and NSF Campus Cyberinfrastructure Grant ACI-1440281.

\bibliographystyle{IEEEtran}

\bibliography{10_bibliography}

\begin{thebibliography}{10}
\providecommand{\url}[1]{#1}
\csname url@samestyle\endcsname
\providecommand{\newblock}{\relax}
\providecommand{\bibinfo}[2]{#2}
\providecommand{\BIBentrySTDinterwordspacing}{\spaceskip=0pt\relax}
\providecommand{\BIBentryALTinterwordstretchfactor}{4}
\providecommand{\BIBentryALTinterwordspacing}{\spaceskip=\fontdimen2\font plus
\BIBentryALTinterwordstretchfactor\fontdimen3\font minus
  \fontdimen4\font\relax}
\providecommand{\BIBforeignlanguage}[2]{{%
\expandafter\ifx\csname l@#1\endcsname\relax
\typeout{** WARNING: IEEEtran.bst: No hyphenation pattern has been}%
\typeout{** loaded for the language `#1'. Using the pattern for}%
\typeout{** the default language instead.}%
\else
\language=\csname l@#1\endcsname
\fi
#2}}
\providecommand{\BIBdecl}{\relax}
\BIBdecl

\bibitem{chang2008bigtable}
F.~Chang, J.~Dean, S.~Ghemawat, W.~C. Hsieh, D.~A. Wallach, M.~Burrows,
  T.~Chandra, A.~Fikes, and R.~E. Gruber, ``Bigtable: A distributed storage
  system for structured data,'' \emph{ACM Transactions on Computer Systems
  (TOCS)}, vol.~26, no.~2, p.~4, 2008.

\bibitem{gadepally2017static}
V.~Gadepally, M.~Hurley, M.~Jones, E.~Kao, S.~Mohindra, P.~Monticciolo,
  A.~Reuther, S.~Smith, W.~Song, D.~Staheli, and J.~Kepner, ``Static graph
  challenge: Subgraph isomorphism,'' in \emph{High Performance Extreme
  Computing (HPEC)}.\hskip 1em plus 0.5em minus 0.4em\relax IEEE, 2017, pp.
  1--6.

\bibitem{gadepally2015gabb}
V.~Gadepally, J.~Bolewski, D.~Hook, D.~Hutchison, B.~Miller, and J.~Kepner,
  ``Graphulo: Linear algebra graph kernels for {NoSQL} databases,'' in
  \emph{International Parallel \& Distributed Processing Symposium Workshops
  (IPDPSW)}.\hskip 1em plus 0.5em minus 0.4em\relax IEEE, 2015.

\bibitem{weale2016benchmarkinggraphulo}
T.~Weale, V.~Gadepally, D.~Hutchison, and J.~Kepner, ``Benchmarking the
  {G}raphulo processing framework,'' in \emph{High Performance Extreme
  Computing (HPEC)}.\hskip 1em plus 0.5em minus 0.4em\relax IEEE, 2016.

\bibitem{hutchison2015graphulo}
D.~Hutchison, J.~Kepner, V.~Gadepally, and A.~Fuchs, ``Graphulo implementation
  of server-side sparse matrix multiply in the {Accumulo} database,'' in
  \emph{High Performance Extreme Computing (HPEC)}.\hskip 1em plus 0.5em minus
  0.4em\relax IEEE, 2015.

\bibitem{bader2014graph}
D.~Bader, A.~Bulu{\c{c}}, J.~Gilbert, J.~Gonzalez, J.~Kepner, and T.~Mattson,
  ``The graph blas effort and its implications for exascale,'' in \emph{SIAM
  Workshop on Exascale Applied Mathematics Challenges and Opportunities
  (EX14)}, 2014.

\bibitem{hutchison2016graphuloalg}
D.~Hutchison, J.~Kepner, V.~Gadepally, and B.~Howe, ``From {NoSQL} {A}ccumulo
  to {NewSQL} {G}raphulo: Design and utility of graph algorithms inside a
  {BigTable} database,'' in \emph{High Performance Extreme Computing
  (HPEC)}.\hskip 1em plus 0.5em minus 0.4em\relax IEEE, 2016.

\bibitem{pavan2013counting}
A.~Pavan, K.~Tangwongsan, S.~Tirthapura, and K.-L. Wu, ``Counting and sampling
  triangles from a graph stream,'' \emph{Proceedings of the VLDB Endowment},
  vol.~6, no.~14, pp. 1870--1881, 2013.

\bibitem{burkhardt2015cloud}
P.~Burkhardt and C.~A. Waring, ``A cloud-based approach to big graphs,'' in
  \emph{High Performance Extreme Computing (HPEC)}.\hskip 1em plus 0.5em minus
  0.4em\relax IEEE, 2015.

\bibitem{bader2006designing}
D.~Bader, K.~Madduri, J.~Gilbert, V.~Shah, J.~Kepner, T.~Meuse, and
  A.~Krishnamurthy, ``Designing scalable synthetic compact applications for
  benchmarking high productivity computing systems,'' \emph{Cyberinfrastructure
  Technology Watch}, vol.~2, pp. 1--10, 2006.

\bibitem{cohen2009graph}
J.~Cohen, ``Graph twiddling in a mapreduce world,'' \emph{Computing in Science
  \& Engineering}, vol.~11, no.~4, pp. 29--41, 2009.

\bibitem{cohen2006user}
S.~Cohen, ``User-defined aggregate functions: bridging theory and practice,''
  in \emph{International Conference on Management of Data (SIGMOD)}.\hskip 1em
  plus 0.5em minus 0.4em\relax ACM, 2006, pp. 49--60.

\bibitem{wolf2015task}
M.~M. Wolf, J.~W. Berry, and D.~T. Stark, ``A task-based linear algebra
  building blocks approach for scalable graph analytics,'' in \emph{High
  Performance Extreme Computing (HPEC)}.\hskip 1em plus 0.5em minus 0.4em\relax
  IEEE, 2015.

\bibitem{leskovec2005realistic}
J.~Leskovec, D.~Chakrabarti, J.~Kleinberg, and C.~Faloutsos, ``Realistic,
  mathematically tractable graph generation and evolution, using kronecker
  multiplication,'' in \emph{PKDD}, vol.~5.\hskip 1em plus 0.5em minus
  0.4em\relax Springer, 2005, pp. 133--145.

\bibitem{gadepally2015using}
V.~Gadepally and J.~Kepner, ``Using a power law distribution to describe big
  data,'' in \emph{High Performance Extreme Computing (HPEC)}.\hskip 1em plus
  0.5em minus 0.4em\relax IEEE, 2015.

\bibitem{kepner2012dynamic}
J.~Kepner, W.~Arcand, W.~Bergeron, N.~Bliss, R.~Bond, C.~Byun, G.~Condon,
  K.~Gregson, M.~Hubbell, J.~Kurz \emph{et~al.}, ``Dynamic distributed
  dimensional data model ({D4M}) database and computation system,'' in
  \emph{International Conference on Acoustics, Speech and Signal Processing
  (ICASSP)}.\hskip 1em plus 0.5em minus 0.4em\relax IEEE, 2012, pp. 5349--5352.

\bibitem{zhai2011cloud}
Y.~Zhai, M.~Liu, J.~Zhai, X.~Ma, and W.~Chen, ``Cloud versus in-house cluster:
  evaluating amazon cluster compute instances for running mpi applications,''
  in \emph{State of the Practice Reports}.\hskip 1em plus 0.5em minus
  0.4em\relax ACM, 2011, p.~11.

\bibitem{azad2016exploiting}
A.~Azad, G.~Ballard, A.~Bulu{\c{c}}, J.~Demmel, L.~Grigori, O.~Schwartz,
  S.~Toledo, and S.~Williams, ``Exploiting multiple levels of parallelism in
  sparse matrix-matrix multiplication,'' \emph{SIAM Journal on Scientific
  Computing}, vol.~38, no.~6, pp. C624--C651, 2016.

\bibitem{beame2014skew}
P.~Beame, P.~Koutris, and D.~Suciu, ``Skew in parallel query processing,'' in
  \emph{Proceedings of the 33rd ACM SIGMOD-SIGACT-SIGART symposium on
  Principles of database systems}.\hskip 1em plus 0.5em minus 0.4em\relax ACM,
  2014, pp. 212--223.

\bibitem{ballard2016hypergraph}
G.~Ballard, A.~Druinsky, N.~Knight, and O.~Schwartz, ``Hypergraph partitioning
  for sparse matrix-matrix multiplication,'' \emph{ACM Transactions on Parallel
  Computing (TOPC)}, vol.~3, no.~3, p.~18, 2016.

\bibitem{buluc2008challenges}
A.~Bulu{\c{c}} and J.~R. Gilbert, ``Challenges and advances in parallel sparse
  matrix-matrix multiplication,'' in \emph{Parallel Processing, 2008. ICPP'08.
  37th International Conference on}.\hskip 1em plus 0.5em minus 0.4em\relax
  IEEE, 2008, pp. 503--510.

\bibitem{chu2015theory}
S.~Chu, M.~Balazinska, and D.~Suciu, ``From theory to practice: Efficient join
  query evaluation in a parallel database system,'' in \emph{Proceedings of the
  2015 ACM SIGMOD International Conference on Management of Data}.\hskip 1em
  plus 0.5em minus 0.4em\relax ACM, 2015, pp. 63--78.

\bibitem{selinger1979access}
P.~G. Selinger, M.~M. Astrahan, D.~D. Chamberlin, R.~A. Lorie, and T.~G. Price,
  ``Access path selection in a relational database management system,'' in
  \emph{Proceedings of the 1979 ACM SIGMOD international conference on
  Management of data}.\hskip 1em plus 0.5em minus 0.4em\relax ACM, 1979, pp.
  23--34.

\bibitem{koanantakool2016communication}
P.~Koanantakool, A.~Azad, A.~Bulu{\c{c}}, D.~Morozov, S.-Y. Oh, L.~Oliker, and
  K.~Yelick, ``Communication-avoiding parallel sparse-dense matrix-matrix
  multiplication,'' in \emph{Parallel and Distributed Processing Symposium,
  2016 IEEE International}.\hskip 1em plus 0.5em minus 0.4em\relax IEEE, 2016,
  pp. 842--853.

\bibitem{shun2015multicore}
J.~Shun and K.~Tangwongsan, ``Multicore triangle computations without tuning,''
  in \emph{International Conference on Data Engineering (ICDE)}.\hskip 1em plus
  0.5em minus 0.4em\relax IEEE, 2015, pp. 149--160.

\bibitem{wang2016myriaOverview}
\BIBentryALTinterwordspacing
J.~Wang, T.~Baker, M.~Balazinska, D.~Halperin, B.~Haynes, B.~Howe,
  D.~Hutchison, S.~Jain, R.~Maas, P.~Mehta, D.~Moritz, B.~Myers, J.~Ortiz,
  D.~Suciu, A.~Whitaker, and S.~Xu, ``The {Myria} big data management and
  analytics system and cloud service,'' in \emph{Conference on Innovative Data
  Systems Research (CIDR)}, 2017. [Online]. Available:
  \url{https://homes.cs.washington.edu/~magda/papers/wang-cidr17.pdf}
\BIBentrySTDinterwordspacing

\bibitem{gadepally2016bigdawg}
V.~Gadepally, P.~Chen, J.~Duggan, A.~Elmore, B.~Haynes, J.~Kepner, S.~Madden,
  T.~Mattson, and M.~Stonebraker, ``The bigdawg polystore system and
  architecture,'' in \emph{High Performance Extreme Computing (HPEC)}.\hskip
  1em plus 0.5em minus 0.4em\relax IEEE, 2016, pp. 1--6.

\bibitem{agrawal2016rheem}
D.~Agrawal, L.~Ba, L.~Berti-Equille, S.~Chawla, A.~Elmagarmid, H.~Hammady,
  Y.~Idris, Z.~Kaoudi, Z.~Khayyat, S.~Kruse \emph{et~al.}, ``Rheem: Enabling
  multi-platform task execution,'' in \emph{Proceedings of the 2016
  International Conference on Management of Data}.\hskip 1em plus 0.5em minus
  0.4em\relax ACM, 2016, pp. 2069--2072.

\bibitem{haynes2016pipegen}
B.~Haynes, A.~Cheung, and M.~Balazinska, ``Pipegen: Data pipe generator for
  hybrid analytics,'' in \emph{Proceedings of the Seventh Symposium on Cloud
  Computing}.\hskip 1em plus 0.5em minus 0.4em\relax ACM, 2016.

\bibitem{dziedzic2016data}
A.~Dziedzic, A.~J. Elmore, and M.~Stonebraker, ``Data transformation and
  migration in polystores,'' in \emph{High Performance Extreme Computing
  (HPEC)}.\hskip 1em plus 0.5em minus 0.4em\relax IEEE, 2016, pp. 1--6.

\bibitem{boehm2014systemml}
M.~Boehm, D.~R. Burdick, A.~V. Evfimievski, B.~Reinwald, F.~R. Reiss, P.~Sen,
  S.~Tatikonda, and Y.~Tian, ``{SystemML}'s optimizer: Plan generation for
  large-scale machine learning programs.'' \emph{IEEE Data Eng. Bull.},
  vol.~37, no.~3, pp. 52--62, 2014.

\bibitem{hutchison2017laradb}
D.~Hutchison, B.~Howe, and D.~Suciu, ``Lara{DB}: A minimalist kernel for linear
  and relational algebra computation,'' in \emph{SIGMOD Workshop on Algorithms
  and Systems for MapReduce and Beyond (BeyondMR)}.\hskip 1em plus 0.5em minus
  0.4em\relax ACM, 2017.

\end{thebibliography}

\balance

\end{document}